\documentstyle[psfig]{mn}

\newif\ifAMStwofonts

\ifoldfss
  \ifCUPmtlplainloaded \else
    \NewTextAlphabet{textbfit} {cmbxti10} {}
    \NewTextAlphabet{textbfss} {cmssbx10} {}
    \NewMathAlphabet{mathbfit} {cmbxti10} {} 
    \NewMathAlphabet{mathbfss} {cmssbx10} {} 
  \fi
  \ifAMStwofonts
    \ifCUPmtlplainloaded \else
      \NewSymbolFont{upmath} {eurm10}
      \NewSymbolFont{AMSa} {msam10}
      \NewMathSymbol{\upi}     {0}{upmath}{19}
      \NewMathSymbol{\umu}     {0}{upmath}{16}
      \NewMathSymbol{\upartial}{0}{upmath}{40}
      \NewMathSymbol{\leqslant}{3}{AMSa}{36}
      \NewMathSymbol{\geqslant}{3}{AMSa}{3E}

    \fi
  \fi
\fi 

\ifnfssone
  \newmathalphabet{\mathit}
  \addtoversion{normal}{\mathit}{cmr}{m}{it}
  \addtoversion{bold}{\mathit}{cmr}{bx}{it}
  \newmathalphabet{\mathbfit} 
  \addtoversion{normal}{\mathbfit}{cmr}{bx}{it}
  \addtoversion{bold}{\mathbfit}{cmr}{bx}{it}
  \newmathalphabet{\mathbfss} 
  \addtoversion{normal}{\mathbfss}{cmss}{bx}{n}
  \addtoversion{bold}{\mathbfss}{cmss}{bx}{n}
  \ifAMStwofonts
    \ifCUPmtlplainloaded \else
      \UseAMStwoboldmath
      \makeatletter
      \new@mathgroup\upmath@group
      \define@mathgroup\mv@normal\upmath@group{eur}{m}{n}
      \define@mathgroup\mv@bold\upmath@group{eur}{b}{n}
      \edef\UPM{\hexnumber\upmath@group}
      \new@mathgroup\amsa@group
      \define@mathgroup\mv@normal\amsa@group{msa}{m}{n}
      \define@mathgroup\mv@bold\amsa@group{msa}{m}{n}
      \edef\AMSa{\hexnumber\amsa@group}
      \makeatother
      \mathchardef\upi="0\UPM19
      \mathchardef\umu="0\UPM16
      \mathchardef\upartial="0\UPM40
      \mathchardef\leqslant="3\AMSa36
      \mathchardef\geqslant="3\AMSa3E
    \fi
  \fi
\fi 

\ifnfsstwo
  \DeclareMathAlphabet{\mathbfit}{OT1}{cmr}{bx}{it}
  \SetMathAlphabet\mathbfit{bold}{OT1}{cmr}{bx}{it}
  \DeclareMathAlphabet{\mathbfss}{OT1}{cmss}{bx}{n}
  \SetMathAlphabet\mathbfss{bold}{OT1}{cmss}{bx}{n}
  \ifAMStwofonts
    \ifCUPmtlplainloaded \else
      \DeclareSymbolFont{UPM}{U}{eur}{m}{n}
      \SetSymbolFont{UPM}{bold}{U}{eur}{b}{n}
      \DeclareSymbolFont{AMSa}{U}{msa}{m}{n}
      \DeclareMathSymbol{\upi}{0}{UPM}{"19}
      \DeclareMathSymbol{\umu}{0}{UPM}{"16}
      \DeclareMathSymbol{\upartial}{0}{UPM}{"40}
      \DeclareMathSymbol{\leqslant}{3}{AMSa}{"36}
      \DeclareMathSymbol{\geqslant}{3}{AMSa}{"3E}
    \fi
  \fi
\fi 

\ifCUPmtlplainloaded \else
  \ifAMStwofonts \else 
    \def\upi{\pi}
    \def\umu{\mu}
    \def\upartial{\partial}
  \fi
\fi

\title{GMRT Observations of NGC~3079}
\author[J. A. Irwin and D. J. Saikia]
       {Judith A. Irwin$^1$ and D. J. Saikia$^2$\\
   $1$ Dept. of Physics, Queen's University, Kingston, Canada, K7L 3N6\\
   $2$ National Centre for Radio Astrophysics, Tata Institute of
Fundamental Research, Pune University Campus, Post Bag 3,\\ 
Pune 411 007,
India}
\date{Accepted.    Received }

\pagerange{\pageref{firstpage}--\pageref{lastpage}}
\pubyear{  }

\begin{document}

\maketitle

\label{firstpage}

\begin{abstract}
We present new observations at three frequencies
(326 MHz, 615 MHz, and 1281 MHz) of the radio lobe spiral galaxy,
NGC~3079, using the Giant Metrewave Radio 
Telescope.  These observations are consistent with
previous data obtained at other telescopes and reveal the structure of
the nuclear radio lobes in exquisite detail.  In addition, new
features are observed, some with HI counterparts, showing broad
scale radio continuum emission and extensions.  
The galaxy is surrounded by a
radio halo that is at least 4.8 kpc in height.  Two giant radio
extensions/loops are seen on either side of the galaxy out to
$\sim$ 11 kpc from the major axis, only slightly offset from the 
direction of the smaller nuclear radio 
lobes.  If these are associated with the nuclear outflow, then
the galaxy has experienced episodic nuclear activity.   Emission along
the southern major axis suggests motion through a local IGM 
(not yet detected) and it may
be that NGC~3079 is itself creating this local intergalactic gas
via outflows.
We also present maps of the minimum energy parameters for this
galaxy, including cosmic ray energy density, electron diffusion length,
magnetic field strength, particle lifetime, and power.
\end{abstract}


\section{Introduction}

NGC~3079 is unusual amongst spiral galaxies in that it displays
two well-defined radio lobes extending $\sim$ 30$^{\prime\prime}$ (2.4 kpc, 
assuming a distance of 16.5 Mpc as listed in Table 1) from the
nucleus (Duric et al. 1983), which
are clearly out of the plane of this edge-on (84$^\circ$) system.
Within the eastern radio lobe is a ``bubble'' of H$\alpha$ emission
extending to 550 pc and displaying
outflowing velocities up to 1000 km s$^{-1}$, as
shown by Hubble Space Telescope (HST) observations (Cecil et al. 2001
and references therein).  Recent Chandra spacecraft images
have also revealed X-ray emitting gas that is spatially
correlated with the emission line gas (Cecil et al. 2002).

The high IR luminosity (Table 1) and
bubble/cone morphology of the emission line gas has led to models of wind-like outflow
related to a starburst.  However, the galaxy also harbours an 
active galactic nucleus (AGN) and jet and about half
 of the IR luminosity could be due to recycled AGN (rather
than hot stellar) photons (Iyomoto et al. 2001). 
The core/jet nature of
the nuclear region is confirmed by VLBI detections (Irwin \& Seaquist 1988; 
Trotter et al. 1998) as well as the recent measurement of an
increase in separation between the compact components implying a jet
outflow velocity of 
 0.13 $\pm$ 0.03 c (Kondratko et al. 2002).
 H$_2$O maser emission in the disk indicates that a
core mass of 10$^6$ M$_\odot$ is present within
a radius of 1 pc.   Recent BeppoSax observations 
(Iyomoto et al. 2001) show that X-ray emission from the core
is highly obscured and that after correction for absorption,
the luminosity is 10$^{42}$ to 10$^{43}$ ergs s$^{-1}$.
In the inner parsec, this is
1\% to 10\% of the Eddington luminosity.
 NGC~3079 is therefore
 as active as the Seyfert galaxies and quasars. While
it is still unclear to what extent the radio core
powers the various components of the kpc-scale outflow,
 the radio lobes themselves appear
to be a nearby analogue of the distant, powerful,
extra-galactic radio sources, but within the dense, high angular
momentum ISM of a nearby spiral. 

Given the strength of the radio continuum emission
in NGC~3079 and its peculiar nature, this galaxy represented a good target for 
low frequency 
observations using the Giant Metrewave Radio Telescope
(GMRT) during its commissioning phase.  We were particularly 
interested in searching for
broad scale emission, especially as might be related to the
nuclear outflow or to some of the other neutral hydrogen
(HI) extensions and supershells seen in this galaxy (Irwin \& Seaquist
1990).  Multi-frequency observations were also important as
a means to test for consistency and to enable the determination
of minimum energy parameters such as magnetic field strength,
cosmic ray energy density, and particle lifetimes.  These are
presented in Sec. 3. The basic data for NGC~3079 are presented in Table 1.

\begin{table*}
 \centering
 \begin{minipage}{140mm}
  \caption{Basic Data on NGC~3079.$^a$}
  \begin{tabular}{@{}cccccccc@{}}
\hline
 RA (J2000)$^b$
& DEC (J2000)$^b$ & Type$^c$ &
 a $\times$ b$^d$ & V$_{sys}$$^e$ & i$^f$ & L$_{FIR}$$^g$ & D$^h$ \\
 (h m s) & ($^\circ$ $^{\prime}$ $^{\prime\prime}$) &
& ($^\prime$ $\times$
$^\prime$) & (km s$^{-1}$) & ($^\circ$) & (L$_\odot$) & (Mpc) \\
\hline
 10 01 57.798 & +55 40 47.08 & SB(s)c &
7.9 $\times$ 1.4 & 1125 & 84 & 4.1 $\times$ 10$^{10}$ & 16.5 \\
\hline
\end{tabular}\hfill\break
$a$ Taken from the Nasa Extragalactic Database, unless otherwise
noted.\hfill\break
$b$ Optical centre coordinates.\hfill\break 
$c$ Morphological type.\hfill\break
$d$ Optical major and minor axes sizes.\hfill\break
$e$ Heliocentric systemic velocity.\hfill\break
$f$ Inclination, from Irwin \& Seaquist (1991).\hfill\break
$g$ Far infra-red luminosity, from the FIR flux listed in
Niklas et al. (1995).\hfill\break
$h$ Distance, from Baan \& Irwin (1995), using H$_0$ = 75 km s$^{-1}$
Mpc$^{-1}$.
\end{minipage}
\end{table*}

\section[]{Observations and Data Reduction}

The GMRT consists of 30 45-m antennas in an approximate `Y'
shape similar to the Very Large Array but
each antenna cannot be moved from a fixed position. 
Twelve antennas are randomly placed within a central
1 km by 1 km square (the ``Central Square'') and the
remainder form the irregularly shaped Y
(6 on each arm) over a total extent of about 25 km.  
Further details about the array can be found at the GMRT website
at {\tt http://www.gmrt.ncra.tifr.res.in}. We present
 details of the observations in Table~2.  
The observations were made in the standard fashion, with
each source observation interspersed with observations
of the phase calibrator.
Data are collected in spectral-line mode at the GMRT
over 128 channels in 2 Stokes (RR and LL) and one IF.
The data are first written
in a GMRT-based ``lta'' format and then converted into
 FITS format to facilitate processing in the
standard fashion using
the Astronomical Image Processing System (AIPS) of
the National Radio Astronomy Observatory (NRAO).

Because no on-line flagging occurred in
real time during GMRT observations, the data were 
first ``pre-edited'' using
the program, GMRED, written by one of us (JAI)
to run within AIPS.  
The purpose of GMRED is to search through each channel,
each stokes and each IF separately and edit out bad data
prior to normal editing and calibration.  It combines
clipping (flagging by a maximum and minimum) as well as median
filtering to eliminate gross outliers. 
The maximum and minimum, the time interval for determination of the 
median, and the deviation allowed from the median before
flagging can all be set by the user after inspection of the raw
UV data.  This allowed for
removal of points such as might be obtained when an
antenna is receiving data but is not yet pointed exactly on
source, points that might be poor at the beginning of scans,
and spikes from interference or electronics. 
GMRED is freely available with instructions for installation
from {\tt http://www.astro.queensu.ca/$\,\tilde{\,}$irwin}. 

We then carried out a calibration in time in the standard
fashion but using a single good channel of
the phase calibrator rather than the inner
75\% of the band.  This was possible since the calibrator
sources were quite strong at each frequency and avoided
potential variations with frequency.

Since the data were acquired in spectral-line mode, 
we also carried out a bandpass
calibration using the phase calibrator.
The data from a
single good channel were then Fourier Transformed
and cleaned, first using a wide field to identify all sources
and then of smaller multiple fields so that the outlying
sources could be cleaned with a beam for the respective positions.
All channels were then self-calibrated
several times (and again edited, if
required) using the single channel map.  
Cleaned maps were then made of all channels (i.e. a cube
was created) and plots of the source flux density 
and map rms noise as a function of channel number were made. 
Loss of signal
at the ends of the channels and occasional interference spikes
or hardware problems required further editing of a number of channels
at this stage.  The final edited
 bandwidths and band centres are given in Table 2. 

We then checked for positional
offsets and variations in source flux density across the band.  We found
that the positions of NGC~3079 (central peak)
 and other point sources within the
field agreed in position from one edge of the band to the other
to within a single pixel (about 1/4 of a beam).  
The standard deviation in the flux density of
NGC~3079 across the band (measuring all channels)
in comparison to the flux density is given in Table 2.  These errors
can be compared to the bootstrapped flux density error on the phase
calibrator as well as the estimated absolute error on
the flux density calibrators (the latter, a few percent)
to arrive at an estimate of
the final calibration error listed in Table 2.

The UV data were then averaged over all good channels to 
make a single-channel UV data set and various maps were
made and cleaned from these data, using a variety of UV
weighting functions and tapers.  Several maps experienced
small improvements from further minor editing
and self-calibration at this stage.
We then corrected for the primary beam at the various
frequencies for all maps.  However, after these corrections,
the increase in source flux density was much less than the
absolute calibration error and the point-by-point change
in flux density per beam was less than the rms map
noise for all maps.  Thus we present the final maps without
this correction.  

We then checked and made minor corrections to
the map registration by comparing the 
positions of the core of NGC~3079 and point sources 
around it (via gaussian fits) 
with those of the
NRAO VLA Sky Survey (NVSS) and
Faint Images of the Radio Sky at Twenty-centimeters
 (FIRST) maps (see Figs. 4b and 4c, respectively).
We estimate the final positional accuracy to be within
 1 arcsec.
We also compared our 326 MHz flux densities with those measured from
the Westerbork Northern Sky Survey (WENSS, Rengelink et al. 1997) map
(Fig. 4a).  
After smoothing to the same resolution, the WENSS map has a peak 
and integrated flux density that are 10\% lower than
our GMRT flux densities.  Given the 10\% accuracy of our GMRT 326 MHz
flux density and the 5\% accuracy of the WENSS flux density, these values
are in agreement. 

\begin{table*}
 \centering
 \begin{minipage}{140mm}
  \caption{Observing and Data Reduction Details.}
  \begin{tabular}{@{}lccc@{}}
\hline
 Observing Parameters & 326 MHz & 615 MHz & 1281 MHz \\
\hline 
 Date & 16 Apr 1999 & 14 Nov 1999 & 03 Dec 2001 \\
 Time$^a$ (h) & 6.2 & 12.3 & 9.0 \\
 Number of antennas$^b$  & 18 & 17 & 26 \\
 Shortest Baseline (k$\lambda$) & 0.0684 & 0.159 & 0.195 \\
 Longest Baseline (k$\lambda$) & 21.8 & 30.9 & 102 \\
 Observing Bandwidth (MHz) & 8 & 16 & 16 \\
 No. of IFs & 1 & 1 & 1 \\
 Flux Density Calibrator(s) & 3C~286 & 3C~48, 3C~286 & 3C~48 \\
\hskip 0.5truein Flux Density (Jy) & 25.94 & 29.38, 21.05   & 17.23\\
 Phase/Bandpass Calibrator  & 0831+557 & 0831+557  & 0831+557\\
\hskip 0.5truein Flux Density$^c$ (Jy) & 8.98 $\pm$ 0.63 & 8.22 $\pm$ 0.13 
& 8.52$\pm$ 0.16\\
 Flux Error across band$^d$ & 9.7\% & 3.3\% & 0.9\% \\
 Final Calibration Error    & 10\%  & 5\%   & 3\% \\
\hline
 Map Parameters & & & \\
\hline
 Final Bandwidth$^e$ (MHz) & 6.2 & 7.75 & 14.0 \\
 Band Centre (MHz) & 325.625  & 615.234   & 1280.700  \\
 Resolution (low) & 59.16 $\times$ 49.56 @ $-$24.7$^\circ$  & 
39.79 $\times$ 29.78 @ 11.3$^\circ$ & 
23.55 $\times$ 19.75 @ 52.1$^\circ$\\
~~~~~~~~~Taper/weighting &      6 k$\lambda$/NA & 
10k$\lambda$/NA   &  10k$\lambda$/NA  \\
~~~~~~~~~rms map noise (mJy beam$^{-1}$) & 4.9 & 1.0 & 0.63 \\
 Resolution (medium) & 44.08 $\times$ 35.53 @ $-$19.8$^\circ$ &
27.08 $\times$  19.65 @ 7.7$^\circ$  &
7.33 $\times$  5.21 @ 48.1$^\circ$ \\
~~~~~~~~~Taper/weighting & none/NA     & none/NA  &  50k$\lambda$/NA   \\
~~~~~~~~~rms map noise (mJy beam$^{-1}$) & 3.4 & 0.9 & 0.25 \\
 Resolution (high) & 28.45 $\times$ 17.66 @ $-$18.2$^\circ$ &
14.90 $\times$ 7.98 @ 36.5$^\circ$ & 
3.22 $\times$ 2.24 @ 38.2$^\circ$\\
~~~~~~~~~Taper/weighting &  none/UN    & none/UN  &  none/UN   \\
~~~~~~~~~rms map noise (mJy beam$^{-1}$) & 2.1 & 0.74 & 0.12\\ 
\hline
\end{tabular}\hfill\break
$a$ Total observing time, including overheads, before editing.
$b$ Maximum number of antennas 
operational at any time during the observations. 
$c$ Flux density and error from GETJY.
$d$  Standard deviation in flux density of
NGC~3079 from measurements of all channels, divided by the mean flux density
of all channels, expressed as a percentage. 
$e$ After editing.
\end{minipage}
\end{table*}

\begin{figure*}
\psfig{figure=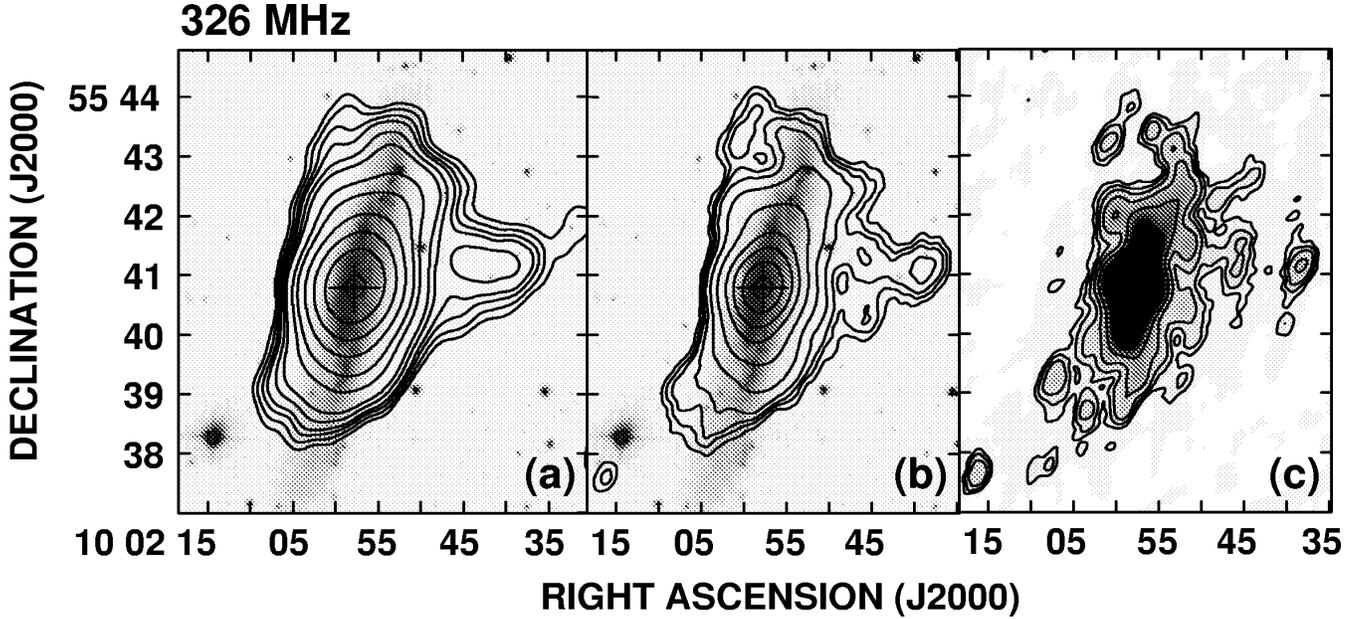,angle=0,width=\textwidth}
\caption{326 MHz maps of NGC~3079 showing different resolutions depending
on UV weighting and tapering (see Table 2).  
The first contour is
at 1.5$\sigma$ in each case and
the NED central position of
NGC~3079 is marked with a cross where the DSS image is shown.
 {\bf (a)} Low-resolution
map showing 326 MHz contours over a DSS optical image. 
Contour levels are at 
7.4, 10.0, 13.0, 16.0,
 22.0, 30.0, 60.0, 100, 200, 400, 800,
and 1600
mJy beam$^{-1}$.  The peak brightness
is 1.73 Jy beam$^{-1}$ and the beam is
59.16$^{\prime\prime}$ $\times$ 49.56$^{\prime\prime}$
at a position angle of $-$24.7$^\circ$. {\bf (b)} 
Medium-resolution
map with contour levels at 
5.1, 7.5, 10.0, 15.0,
30.0, 75.0, 200, 400, 800, and 1300
 mJy beam$^{-1}$ over a DSS image.  
The peak brightness
is 1.46 Jy beam$^{-1}$ and the beam is
44.08$^{\prime\prime}$ $\times$ 35.53$^{\prime\prime}$
at $-$19.8$^\circ$.
{\bf (c)} 
High-resolution
map (greyscale plus contours) with
levels at 
3.1, 5.0, 8.0, 15.0,
 30.0, and 75.0  mJy beam$^{-1}$.  The greyscale ranges
from 0 to 50 mJy beam$^{-1}$.
The peak brightness
is 987 mJy beam$^{-1}$ and the beam is
28.45$^{\prime\prime}$ $\times$ 17.66$^{\prime\prime}$
at $-$18.2$^\circ$.
\label{maps-fig1}
}
\end{figure*}
 
\begin{figure*}
\psfig{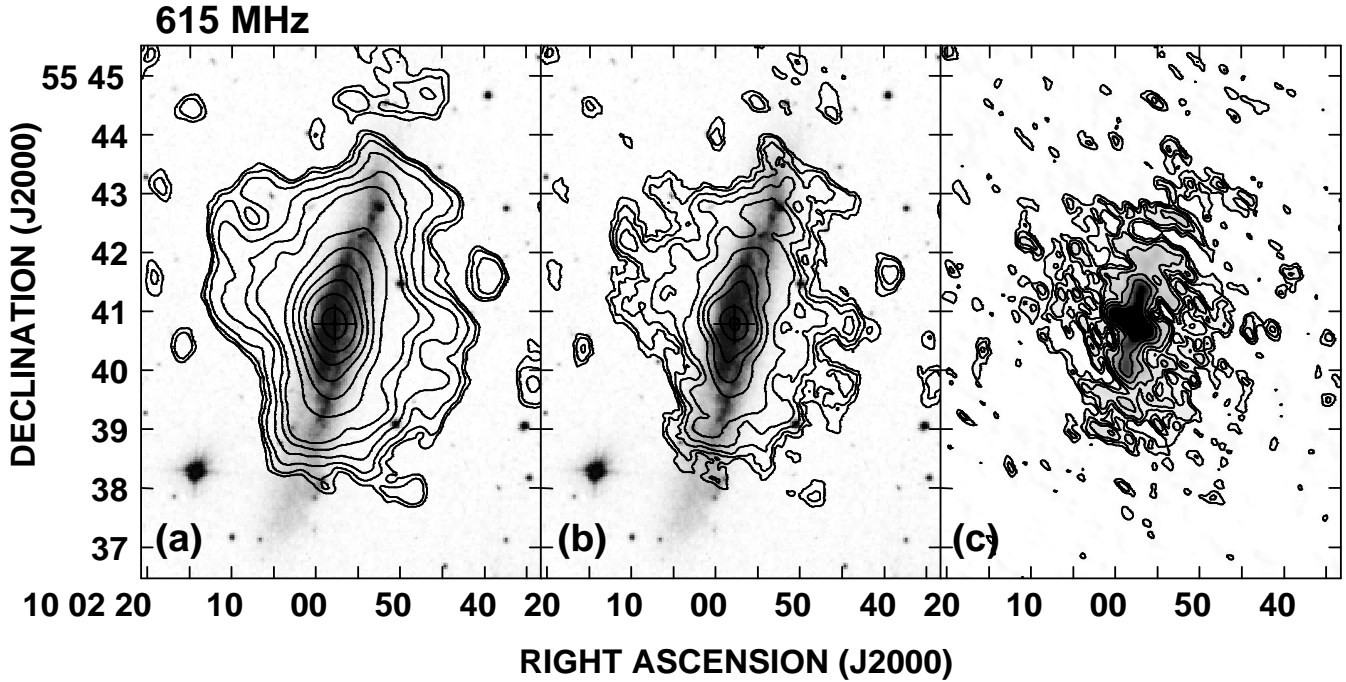}
\caption{The 615 MHz maps of NGC~3079 (contours) over the DSS
image (greyscale) except for (c) for which the greyscale
is the 615 MHz map itself.
Different resolutions are shown depending
on UV weighting and tapering (see Table 2).  The first contour is
at 1.5$\sigma$ in each case and a cross marks the central position
of NGC~3079, in images for which the greyscale is the DSS image.
 {\bf (a)} 
Contour levels are at 
 1.5, 2.0,  3.0,  5.0,
     7.5,   10, 20,    40,
     80,  150, 300,     600,
  830 
 mJy beam$^{-1}$.  The peak brightness
is 832 mJy beam$^{-1}$ and the beam is  
39.79$^{\prime\prime}$ $\times$ 29.78$^{\prime\prime}$
at a position angle of 11.3$^\circ$. {\bf (b)} 
Medium-resolution
map with
contour levels at 
1.3,    2.0,     3.0,      4.5,
      7.0,            18,     40,    80,
300,           600, and 670 
 mJy beam$^{-1}$.  
The peak brightness
is 677  mJy beam$^{-1}$ and the beam is
27.08$^{\prime\prime}$ $\times$ 19.65$^{\prime\prime}$
at 7.7$^\circ$.
{\bf (c)} 
High-resolution
map with
contour levels at 
1.2, 2.0, 3.0, 5.0, 10, 20, and 40
 mJy beam$^{-1}$.  
The peak brightness
is 445 mJy beam$^{-1}$ and the beam is
14.90$^{\prime\prime}$ $\times$ 7.98$^{\prime\prime}$
at 36.5$^\circ$.  
The greyscale ranges from 0 to 35 mJy beam$^{-1}$.
\label{maps-fig2}
}
\end{figure*}

\begin{figure*}
\psfig{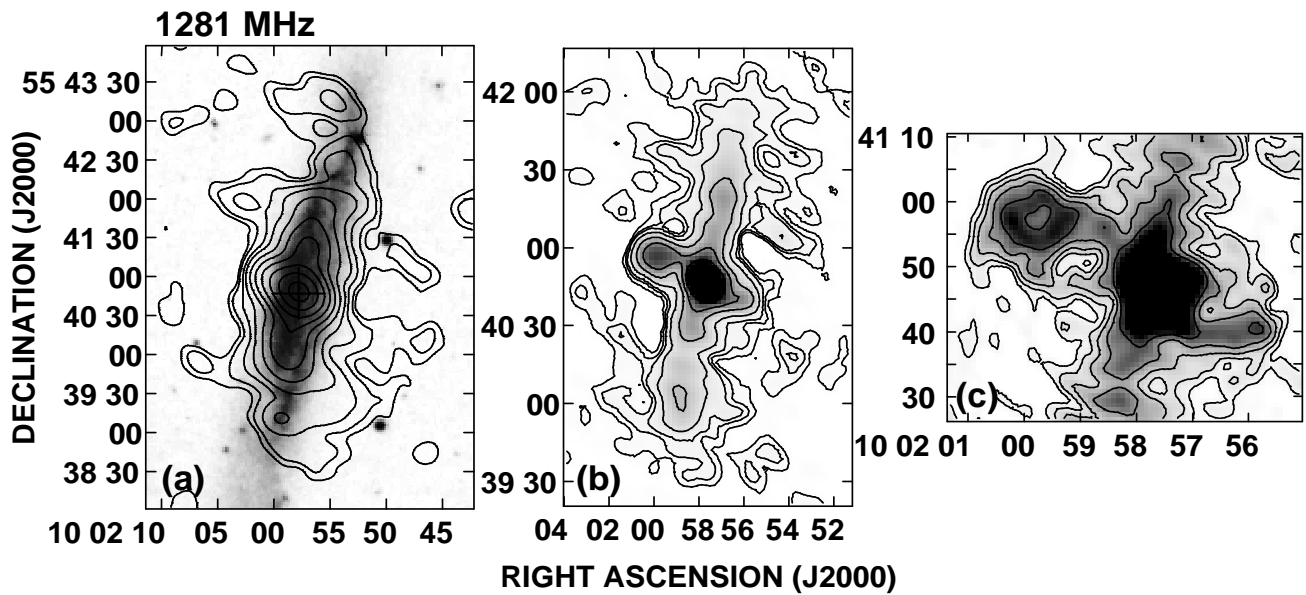}
\caption{The 1281 MHz maps of NGC~3079 (contours).
Different resolutions are shown depending
on UV weighting and tapering (see Table 2).  The first contour is
at 1.5$\sigma$ in each case.
 {\bf (a)} 1281 MHz contours over the DSS image (greyscale).  A 
cross marks the position of the centre of the galaxy.
Contour levels are at 0.9, 1.5, 3.0, 5.0, 10, 
  20,    40,
     80,  160, and 300
 mJy beam$^{-1}$.  The peak brightness
is 384 mJy beam$^{-1}$ and the beam is  
23.55$^{\prime\prime}$ $\times$ 19.75$^{\prime\prime}$
at a position angle of 52.1$^\circ$. {\bf (b)} 
Medium-resolution
map (contours plus greyscale) with
contour levels at 
0.4, 0.8, 1.2, 2.5,
      6.0, 12, and 25  
 mJy beam$^{-1}$.  
The peak brightness
is 201  mJy beam$^{-1}$ and the beam is
7.33$^{\prime\prime}$ $\times$ 5.21$^{\prime\prime}$
at 48.1$^\circ$.  The greyscale
ranges from 0 to 25 mJy beam$^{-1}$.
{\bf (c)} 
High-resolution
map showing the details of the circumnuclear region.  Contours
are at 0.2, 0.5, 1.0, 2.0, and 3.5
 mJy beam$^{-1}$.  
The peak brightness
is 130 mJy beam$^{-1}$ and the beam is
3.22$^{\prime\prime}$ $\times$ 2.24$^{\prime\prime}$
at 38.2$^\circ$.  
The greyscale ranges from 0 to 5 mJy beam$^{-1}$.
\label{maps-fig3}
}
\end{figure*}

\begin{figure*}
\psfig{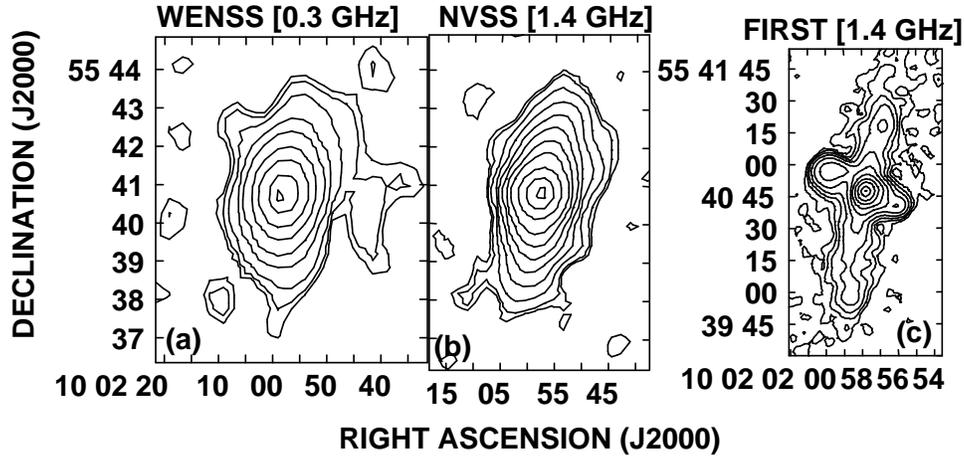}
\caption{
{\bf (a)} WENSS map of NGC~3079 with contour levels at
6.6 (1.5$\sigma$), 10.5, 20, 50, 100, 250, 500, 1000, and 1400
mJy beam$^{-1}$.  The peak brightness is 
4.76 Jy beam$^{-1}$ and the beam is
54.0$^{\prime\prime}$ $\times$ 65.38$^{\prime\prime}$
at a position angle of 0.0$^\circ$.
 {\bf (b)} NVSS map, with
contour levels at 0.72 (1.5$\sigma$), 1.0, 2.0, 4.0, 
 8.0, 15, 30, 60, 150, 300, and 460
 mJy beam$^{-1}$.  The peak brightness
is 472 mJy beam$^{-1}$ and the beam is  
45.0$^{\prime\prime}$ $\times$ 45.0$^{\prime\prime}$
at 0.0$^\circ$. {\bf (c)} 
FIRST map with
contour levels at 
 0.225 (1.5$\sigma$), 0.5, 1.0, 2.0, 4.0, 8.0, 15,
30, 60, 120, and 170
 mJy beam$^{-1}$.  
The peak brightness
is 173  mJy beam$^{-1}$ and the beam is
5.4$^{\prime\prime}$ $\times$ 5.5$^{\prime\prime}$
at 0.0$^\circ$.  There are visible
stripes in this image parallel to the direction of the major
axis.
\label{maps-fig4}
}
\end{figure*}

\begin{figure*}
\psfig{figure=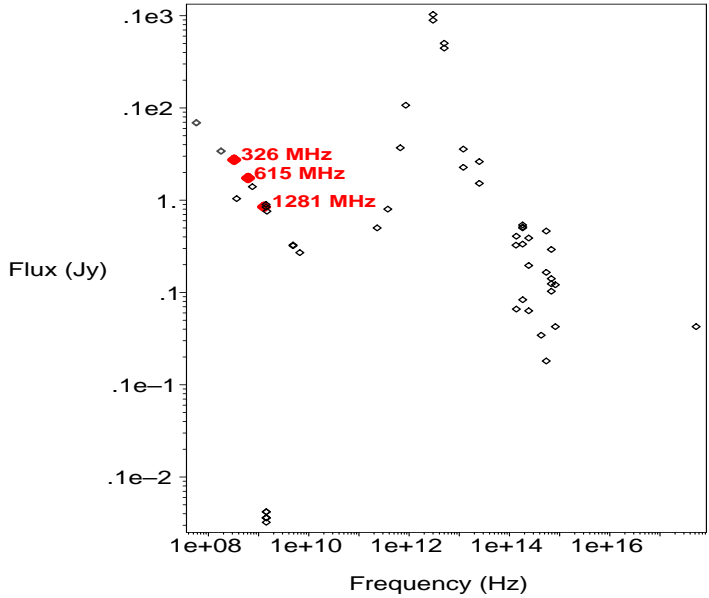,height=8cm,width=12cm}
\caption{Spectrum of NGC~3079, taken from NED, and including the new GMRT flux
density points (shown slightly larger) at 326, 615, and 1281 MHz.
\label{maps-fig5}
}
\end{figure*}

\begin{figure*}
{\hspace{0.1in}
\psfig{figure=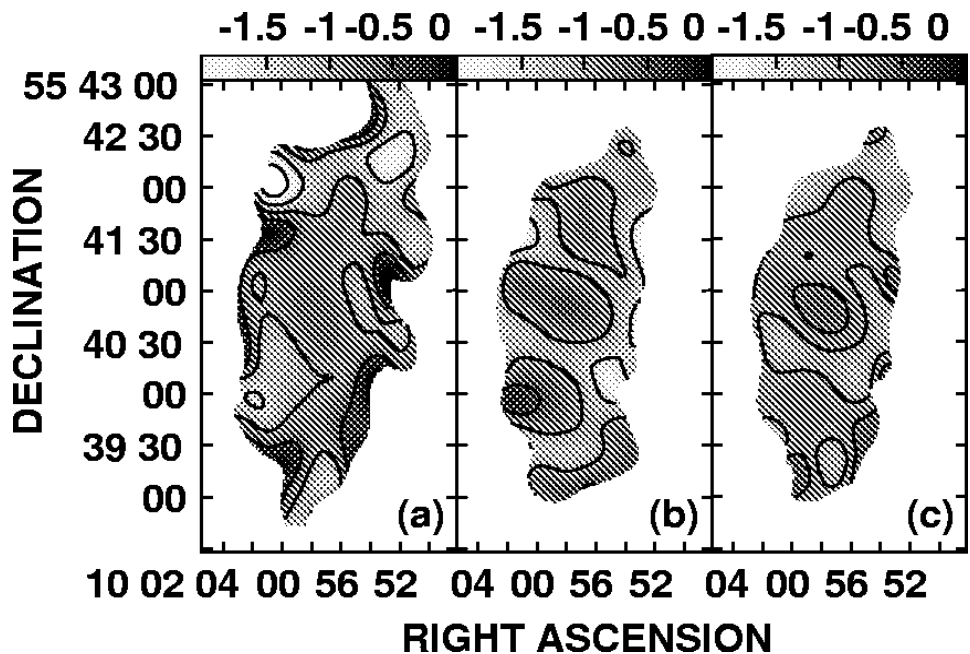,width=5.0in}
}
\caption{
{\bf (a)} Spectral index map between 326 and 615 MHz
($S_\nu\propto\nu^\alpha$).  Contours are at $-$2, 
$-$1.5, $-$1, $-$0.5, and 0.  The greyscale is shown at the top
and so throughout.  The mean of this map is $-$0.96 ($-$0.92 if
restricted to the region shown in c).  A typical error on
any point is 0.10 (see text).
 {\bf (b)} Spectral index map between 615 and 1281 MHz
with contours at $-$1.5, $-$1, and $-$0.5.  The map mean is
$-$1.01 ($-$1.02 if restricted to the region shown in c).  A typical
error on any point is 0.1.
 {\bf (c)} 
Average of maps (a) and (b).  Contours are at
$-$1.0 and $-$0.75.  The mean of the map is $-$0.97 and the
rms is 0.18.
\label{maps-fig6}
}
\end{figure*}

\begin{table*}
 \centering
 \begin{minipage}{140mm}
  \caption{Global Radio Continuum Properties}
  \begin{tabular}{@{}cccccc@{}}
\hline
  & 326 MHz & 326 $\to$ 615
 & 615 MHz & 615 $\to$ 1281 & 1281 MHz \\
\hline
 Flux Density (Jy) & 2.8 $\pm$ 0.3 &  & 1.74 $\pm$ 0.09 & & 
0.85 $\pm$ 0.03\\
Spectral Power (10$^{22}$ W Hz$^{-1}$) & 
9.1 $\pm$ 0.9 & & 5.7 $\pm$ 0.3
& & 2.8 $\pm$ 0.1\\
 Global Spectral Index$^a$ & & $-$0.75 $\pm$ 0.10 & & $-$0.98 $\pm$ 0.04 & \\
 Mean Spectral Index$^b$ & & $-$0.92 $\pm$ 0.04 & & $-$1.02 $\pm$ 0.05 &\\
\hline
\end{tabular}\hfill\break
$a$ Spectral index as determined from the total flux densities given in this
table.  \\
$b$ Mean as determined over a region that is
equal in size between the two spectral index maps, i.e. that
shown in Fig. 5c.  See text for discussion of errors.
\end{minipage}
\end{table*}

\section{Results}

\subsection{Total Intensity Maps}

The total-intensity maps, shown over a variety of spatial
scales, can be seen in Figs. 1 (326 MHz), 2
(615 MHz), and 3 (1281 MHz). Their map properties
are listed in
Table 2.  In these figures, we show positive contours to low intensity
levels (1.5$\sigma$) but caution in the interpretation of
the lowest levels and summarize what we consider to
be real features in
Sect. 3.1.4.
Total flux densities from these maps are
given in Table 3 and are plotted on the spectrum shown in 
Fig. 5.  The new GMRT flux densities are consistent with
the known spectrum at these frequencies.  
(The one discordant
low point at 365 MHz is from the Texas Survey 
using data taken between the years 
 1974 and 1983; see
Douglas et al. 1996.) 
The new data are clearly and firmly in the
non-thermal part of the spectrum.

\subsubsection{326 MHz}

Our low-resolution GMRT image (Fig. 1a)
shows strong emission from NGC~3079, but  
no emission from either of its two companion galaxies, namely NGC~3073 and MCG-9-17-9,
the upper limit to their flux densities being 9 mJy 
and 
5 mJy, respectively. 

At 326 MHz (Fig. 1) an 
envelope of emission is seen around
the galaxy  which
is more extensive than previously observed
(cf. Duric \& Seaquist 1988, Duric et al. 1983,
de Bruyn 1977, and
Fig. 4) with the possible exception of the low-resolution image of
Condon (1987).  The emission envelope follows the
disk of the galaxy and
is likely due to distributed sources in the disk
similar to other star-forming spirals.

To determine the vertical scale height of this thick disk, we
created a model galaxy with the geometry of
NGC~3079, whose projected intensity
at any point is proportional to the line
of sight distance through the galaxy at that point.
The model galaxy was then convolved 
with a beam equivalent
to that of Fig. 1a.  We then matched the extent
of the model major axis with that of NGC~3079 at the 2$\sigma$
level and varied the vertical height until a match was
achieved in the perpendicular direction (the western
extension was omitted, see below).  The best fit
was found for a vertical height of 60$^{\prime\prime}$
(4.8 kpc).  Since the intrinsic brightness distribution
should fall off with height from mid-plane, the vertical
scale height will be larger than 4.8 kpc.

The 326 MHz emission also shows much
substructure with improving resolution. 
The most obvious feature
is an extension towards the west which ends in a knot at
RA = 10 01 38.1, DEC = 55 41 14. A feature is also seen
at a similar position in the  
WENSS image (Fig. 4a) at the same frequency though
with somewhat different structure, given the
different uv coverage and spatial resolution.
This feature resolves into two distinct extensions
(Fig. 1c) which roughly flank the western nuclear
radio lobe.  These two extensions have 
counterparts in HI, best seen in the moment map
of Irwin \& Seaquist (1987).
  
A large loop of emission is visible at
the north end of the galaxy disk extending towards the
east of it.  The north edge of this loop makes the
major axis on the north side appear to bend around
to the east.  
The central minimum is at RA = 10 01 58.3, DEC = 55 43 01
 and the loop diameter
is 47$^{\prime\prime}$
(3.8 kpc)  parallel to the major axis
and extends to 90$^{\prime\prime}$ 
(7.2 kpc) from the major axis.  
This loop is also coincident with an HI (open-topped)
loop seen at a velocity of
1011 km s$^{-1}$ in Irwin \& Seaquist (1991, their Fig. 1).  

Another interesting result is the 
curvature of the emission towards the east
away from the optical disk
at the southern tip of the major axis.  
This behaviour again mimics that
seen in the neutral hydrogen distribution
(Irwin \& Seaquist 1991).  The NVSS image (Fig. 4b)
also hints at an extension in this direction.  
Since both the radio continuum and HI emission are
disconnected from the optical disk, 
the curvature is not tidal but is likely due to
ram pressure affecting only the ISM components.  
This
indicates that there may be more extensive intergalactic gas
in this region through which NGC~3079
has some local motion.  Such motion
 would be (in projection)
towards the west 
or south-west or could be rotation (tumbling) with
the southern major axis moving towards the west.
 At the present time, the only extended 
X-ray emission observed appears to be
associated with the nuclear outflow (Pietsch et al. 1998,
Cecil et al. 2002).
Thus deeper X-ray observations are warranted.

\subsubsection{615 MHz}

The 615 MHz data (Fig. 2) reveal the largest radio continuum loop
which has yet been
associated with this galaxy.  
The loop extends to the east, 
has a central minimum located at
RA = 10 02 07.4, DEC = 55 42 37, has a
diameter of 64$^{\prime\prime}$ (5.1 kpc) parallel
to the major axis and extends to 2.2$^{\prime}$
(11 kpc) from the major axis. 
This is much farther than the minimum vertical
scale height measured from the 326 MHz data and confirms
that a very extensive radio continuum halo exists around
NGC~3079.   The loop is not
centered over the eastern nuclear radio lobe, but
rather offset from it to the north.  However, there is much
substructure in the feature, especially on its
southern edge which is closer to the position of
the radio lobe.  It is possible that several features
are blended along the line of sight in this edge-on
system.  Alternatively, the feature may be related
to the nuclear outflow but 
source precession, motion of the galaxy with
respect to an IGM, or flows into previously
emitted gas may have affected its position
over time.  This feature is also hinted at in the WENSS image
(Fig. 4a).

The same two western extensions on the west side of
the disk (as noted from the 326 MHz data) are seen
prominently at 615 MHz as well.
These (Fig. 2b)
reach $\sim$ 2$^{\prime}$ (9.6 kpc)
from the plane. 

The distinctive cross shape and markedly stronger
emission of the nuclear region
in this galaxy is especially visible in Fig. 2c.

\subsubsection{1281 MHz}

At the highest frequency (Fig. 3), we 
again see emission related to the northern loop at
RA $\sim$ 10 01 55, DEC $\sim$ 55 43 18,
a feature related to the giant eastern loop at
RA $\sim$ 10 02 03, DEC $\sim$ 55 42 00,
and the two western extensions.
At these higher resolutions, we also detect
details of the nuclear radio lobes.
Fig. 3b, for example, 
can be compared to the 1420 MHz FIRST image of Fig.
4c which is quite comparable in structure, except
that our 1281
MHz image has a slightly lower resolution and 
shows some additional emission away from the plane.  (Note that 
the FIRST image contains a noticeable stripe.)
Fig. 3c reveals the lobes' substructure in exquisite
detail, especially the loop-like nature of the
eastern lobe (cf. Duric \& Seaquist 1988).  The
H$\alpha$ emission occurs roughly in the region of
the gap between the eastern loop and the nucleus
(Cecil et al. 2001).

\subsubsection{Summary of Radio Continuum Structure}

Each data set
set has its own limitations, via uv coverage,
spatial resolution, and possible residual errors which
could not be eliminated.  We will therefore
consider a feature to be
real only if it is observed at more than one frequency
(note that all data were taken on different dates,
Table 2) or
if it is observed in another independent observation
using a different telescope.  Thus we summarize the
structures that we believe to be real as follows:

1.  The western extension (Fig. 1a) which resolves into
two distinct extensions (Fig. 1c, Fig. 2a, 2b, Fig.
3a, 3b, Fig. 4a, and is also seen in HI, Irwin \& Seaquist 1987).
These features extend to $\sim$ 10 kpc, 
roughly flank the western nuclear radio
lobe and may be associated with it.

2.  The northern loop (Fig. 1b, c, Fig. 3a and HI,
Irwin \& Seaquist 1991) of 3.8 kpc diameter and extending
7.2 kpc from the major axis.  This loop appears to be
associated with the disk.

3.  The giant eastern radio loop (Fig. 2a, 2b, Fig. 3a,
Fig. 4a) of 5.1 kpc diameter and extending 11 kpc from
the major axis.  It is offset from the eastern nuclear
radio lobe but could still be related to larger scale
and/or earlier nuclear outflow.

4.  The offset, curved emission on the south tip of
the major axis (Fig. 1a, 1b, 1c, Fig. 4b and HI,
Irwin \& Seaquist 1991), suggesting that there may
be motion through broader scale IGM gas.

\subsection{Spectral Index Maps}

To make the spectral index maps, we took Fig. 1c, Fig. 2b,
and Fig. 3a, which are all at similar spatial resolutions,
smoothed them to a common resolution, and blanked all
points below a level of 
5 sigma.  Spectral index maps
($S_\nu\,\propto\,\nu^{\alpha}$)
were then formed and are shown in Fig. 6. Fig. 6a shows the
326-615 MHz spectral index and
Fig. 6b shows the spectral index map between
615 and 1281 MHz. 
To estimate the error in the spectral index maps, 
at each freuency we extracted a map of
noise from a region in which there was no emission.
We then added this noise map to the map of the emission
region of NGC~3079.  This resulted in 3
maps of emission with added noise typical 
of that data set.  New spectral index maps
were then formed from the noise-added images. 
We then formed
a map of the difference between the noise-added spectral
index maps and the original spectral index maps.   The
rms of these difference maps is typical of the error
on any individual point in Figs. 6a and 6b, i.e. $\sim$ 0.1.

In Table 3, we list values of the spectral index, as determined
from the total fluxes listed in Table 3 (the Global spectral index)
and as determined from the mean of the spectral index maps over
equivalent regions (the Mean spectral index).  
Note that the mean spectral index, which applies to
high S/N regions,  shows very little 
change with
changing frequency and the global spectral index is consistent
with these values between 615 and 1281 MHz ($\alpha$ = $-$0.92 to $-$1.02).
However, the 326$-$615 MHz global value is flatter at $-$0.75.
This suggests that there may be more faint extended emission at
326 MHz which remained undetected.  There may indeed be some trend,
however, for stronger
changes in spectral index in the high latitude gas than in the
disk.  For example, extrapolating the 615 MHz emission in the broad
extended halo region
 using $\alpha$ = $-$0.98
to 326 MHz should have revealed as extended a halo
in Fig. 1c above the noise.  A trend towards a flatter spectral
index at lower frequencies can also
be observed in the global spectrum (Fig. 5) when other independent
data points are considered. 
Flattening
of the spectral index at lower frequencies in the non-thermal
domain has been observed before in other galaxies.  Israel
\& Mahoney (1990) attribute such flattening to thermal absorption but
this is unlikely to play a major role in our data since the
mean spectral index, which applies mainly to the galaxy disk where most
of the thermal gas should occur,
 shows only very minor flattening.
Energy losses at the higher frequencies seems more likely
(Hummel 1991).

In this regard, it is interesting to examine the distribution of spectral index
as shown in Fig. 6a and Fig. 6b
for differences.  We note that there are differences
between the maps that are greater than a typical error on any
point.  
For example, a map of the difference between Fig. 6a and 6b
(not shown)
has an rms of 0.49 whereas a typical error bar is about 0.2.  However,
closer examination shows that the larger differences occur
near the edges of the maps (where the S/N is lower) or in regions in which
there are strong intensity gradients (where 1$^{\prime\prime}$
errors in registration can make larger changes in 
the measured value of $\alpha$).  Therefore, in the region shown in
Fig. 6c, we see no convincing evidence for changes in the spectral
index with frequency above the error bars.  
Therefore, we averaged
Fig. 6a and 6b to create a map of the average
spectral index, shown in Fig. 6c.

Fig. 6c has a mean of $\alpha$ = $-$0.97 and an rms of
0.18 with a typical error less than 0.1.  There are
hints that the spectral index may flatten at the bases
of extensions (e.g. on the west side).  The region
centered on the nucleus clearly has a flatter spectral index
than in other regions of the disk,
e.g. a mean of $-$0.73 in the central $\approx$
30$^{\prime\prime}$ diameter region, in agreement with
previous observations of this region (Duric \& Seaquist
1983).     The flatter
values near the nucleus are consistent with the 
fact that this is a region in which cosmic ray acceleration
is occurring and particles are younger.

\subsection{Computation of Minimum Energy Parameters}

From the mean spectral index map (Fig. 6c) and one total intensity map
(the 615 MHz map that was used as to create the images of Fig. 6), we
can compute the following minimum energy parameters
(see Pacholczyk 1970, Duric 1991): the cosmic ray energy density,
$u_{CR}$, the cosmic ray electron 
diffusion length, $L_D$,  magnetic field strength, $B$, and
the particle lifetime, $t$. These quantities
are computed for each pixel
and represent means along a line of sight. 
 A line of sight distance,
and therefore a geometry, is required for these calculations.
We also require
assumptions regarding
the lower and upper frequency cutoffs of the spectrum,
$\nu_1$ and $\nu_2$, and the
ratio of the heavy particle energy to the electron energy, $k$.
(See Irwin et al. 1999 for a previous example.)
Values which have been used are specified via several
models which are listed in Table 4.  In
Fig. 7, we show the results for Model 1 which uses an inclined disk, $k\,=\,40$, 
$\nu_1\,=\,10^7\,\,$Hz and $\nu_2\,=\,10^{11}\,\,$Hz.  Table 4 lists the means 
calculated over these maps as well as those of the other models. 

The results for Models 1 to 4 (Table 4) show 
that changes in geometry (inclined disk or
oblate spheroid) and  changes in the upper frequency cutoff make 
relatively
small differences in the results.  The main differences are due to
the assumption of heavy particle to electron energy (cf. Model 1, Model 2)
which introduce variations in the minimum energy parameters of order
a factor of $\approx$ 1.5. 
Thus, we expect Model 1 (Fig. 7) to be a good representation of
the data to within 
 a factor of $\approx$ 2.
Note that this choice affects only the absolute scale
of the 
maps of Fig. 7 and not the point to point variation. 

The fact that the measured flux densities for NGC~3079 fall within the non-thermal
regime of the spectrum (Fig. 5) indicates that the thermal fraction should be
very small.  However, point to point variation in thermal fraction {\it could}
change the appearance of the minimum energy parameter maps.   
Condon (1992) has provided an estimate (thought to
be correct to within a factor of $\approx$ 2) of the thermal fraction
in spiral galaxies, as measured globally, viz.
 ${S/S_{T}}\, \sim\,  1\,+\,10\big({\nu/GHz}\big)^{(0.1-|\alpha_{NT}|)}$,
where $\alpha_{NT}$ is the non-thermal spectral index, $S$ is the total
flux density and $S_T$ is the flux density due to the thermal component.
Taking
$\alpha_{NT}$ to be $-$0.97 (from the mean spectral index map of
Fig. 6c), we find that only 
3.6\%, 6\%, and 11\%
of the total emission could be due to thermal components
at 326, 615, and 1281 MHz, respectively.  To see what such a contribution
might do to the calculation of the minimum energy parameters, we subtracted
the above fractions from the total intensity images at the 3 frequencies,
recomputed the spectral index map and recomputed the minimum energy
maps.  These results are also listed in Table 4 (Model 5) and show
that there is a negligible difference in the minimum energy parameters
as a result of possible thermal contributions.  

In Fig. 7, we also show a map of power, $P\,=\,U/t$,
where $U$ is the cosmic ray energy density, $u_{CR}$ (Fig. 7a), 
integrated along a line of sight and $t$ is the particle lifetime
(Fig. 7d).  The result is identically,
$P\,=\,(1\,+\,k)\,L$, where $L$ is the observed luminosity at a point.  Thus the
map of $P$ closely resembles the map of total flux density but does not match it exactly 
because the computation of $L$ requires an integration over frequency
which is dependent on spectral index,
 and the spectral index is different
at different points in the map. The map of $P$ 
represents the rate at which cosmic rays
must be accelerated in order to maintain equilibrium.  Integrated over
the map, the total power is 
$P_T\,=\,9.2\times\,10^{41}$\,\,ergs\, s$^{-1}$.  The total
energy in cosmic rays is 
$U_T\,=\,1.1\times\,10^{57}$\,\,ergs.  These values
are minima since not all of the galaxy's emission is represented in the
regions shown in Fig. 7.  In comparison, the values for the Galaxy are
$P_T\,\sim\,2\,\times\,10^{40}$\,\,ergs\,s$^{-1}$ and
$U_T\,\sim\,10^{55}$\,\,ergs (Condon 1992).  Thus the cosmic ray energy generation
rate and total energy are two orders of magnitude higher in NGC~3079
than in the Galaxy.  The spectral power of Table 3 suggests a
supernova rate of 0.27 yr$^{-1}$ (see Condon 1992) if all emission
has this origin.  This is clearly an upper limit since some
of the emission is due to the AGN and associated outflow.

Although the total cosmic ray energy and power  
are higher than the Galaxy, they are not atypical
of spiral galaxies (Duric 1991).  Our average magnetic
field strength of $\sim$ 10 $\mu$Gauss, for example, agrees
with the 12 $\mu$Gauss found for NGC~6946
(Ehle \& Beck 1993).  In addition, the distribution of values
is as expected.
The magnetic field strength, energy
density, and power are all higher near the nucleus where there is
much activity and correspondingly, the lifetime of the particles and
the diffusion length are shorter.  We note, in addition, that
the peak power of $8\,\times\,10^{38}$ ergs s$^{-1}$ is considerably
smaller than the power of the X-ray core in this galaxy
($10^{42-43}$ ergs s$^{-1}$, Sec. 1).

\begin{table*}
 \centering
 \begin{minipage}{140mm}
  \caption{Minimum Energy Parameters}
  \begin{tabular}{@{}lcccc@{}}
\hline
  Model$^a$ & $\overline{U_{CR}}$ & $\overline{L_D}$
 & $\overline B$ & $\overline{t}$ \\
  &  (eV cm$^{-3}$) & (Kpc) & ($\mu$Gauss) & (Myrs) \\
\hline
{\bf (1)} Inclined Disk ($k\,=\,40$, $\nu_2\,=\,10^{11}\,Hz$)
 & 8.03 & 1.38  & 9.32  & 88.1 \\
{\bf (2)} Inclined Disk ($k\,=\,100$, $\nu_2\,=\,10^{11}\,Hz$)
 & 13.4 & 1.07  & 12.1 & 59.1  \\
{\bf (3)} Inclined Disk ($k\,=\,40$, $\nu_2\,=\,10^{10}\,Hz$)
 & 7.82 & 1.40  & 9.21 &  116.3  \\
{\bf (4)} Inclined Oblate Spheroid ($k\,=\,40$, $\nu_2\,=\,10^{11}\,Hz$) 
& 8.42  &  1.34 & 9.58 &  82.4 \\
{\bf (5)}  Inclined Disk ($k\,=\,40$, $\nu_2\,=\,10^{11}\,Hz$, `thermal subtracted')
 & 8.28 & 1.36  & 9.47  & 96.7 \\
\hline
\end{tabular}\hfill\break
$a$ All models adopt a semi-major axis of 
131.5$^{\prime\prime}$
(measured to 5$\sigma$ from the 615 MHz map used to create Fig. 6), 
a semi-minor axis of 
60$^{\prime\prime}$,
an inclination of 84.3$^\circ$ (Irwin \& Seaquist 1991),
and a lower frequency cutoff of $\nu_1\,=\,10^7\,Hz$.  \\
\end{minipage}
\end{table*}

\begin{figure*}
\psfig{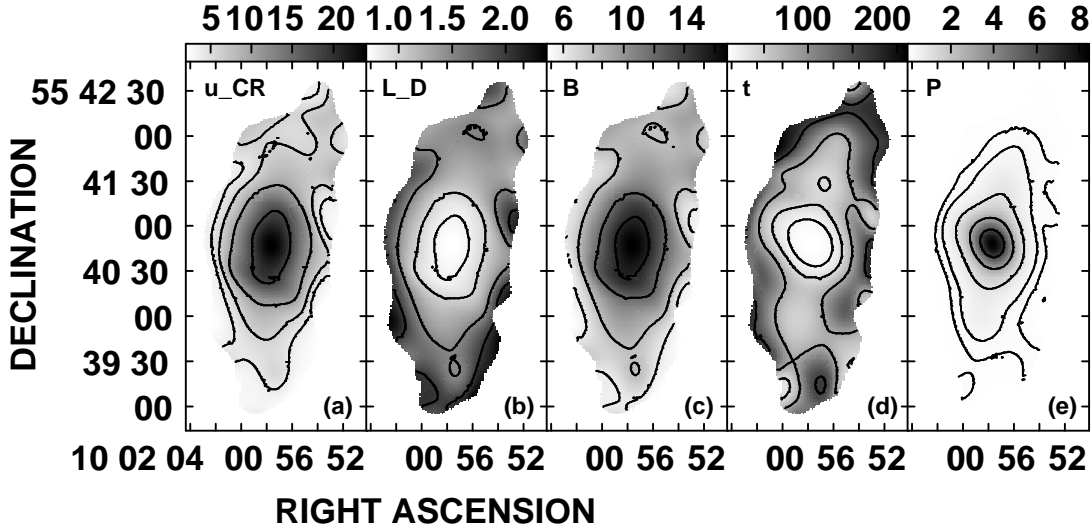}
\caption{Minimum energy values, computed as means over a line of sight
at each pixel (except for e)
corresponding to Model 1 of Table 4.  
The greyscale range is shown at the top and so
throughout.
{\bf (a)} Cosmic ray energy density.  Contours are at 2, 
4, 6, 10, and 18 eV cm$^{-3}$ with minimum and maximum values of 
2.1 and 24.0 
 eV cm$^{-3}$, respectively. The total (cosmic ray plus magnetic field)
energy density would be 7/4 times these values.
 {\bf (b)} Diffusion length.  Contours are at 
0.8, 1, 1.5, and 2 kpc with minimum and maximum values of 0.7 and 2.4 kpc.
 {\bf (c)} 
Magnetic field strength.  Contours are at 6, 8, 12, and
15 $\mu$Gauss with minimum and maximum values of 5.0 and 16.8 $\mu$Gauss.
{\bf (d)} Particle lifetime.  Contours are at 20, 40, 80 and
160 Myrs with minimum and maximum value of 10 and 212 Myr.
{\bf (e)} Map of total power along a line of sight.  Contours are at
0.15, 0.3, 1, 3, and 6 $\times\,10^{38}$ ergs s$^{-1}$ with minimum
and maximum values at 0.04 and 8.4 $\times\,10^{38}$ ergs s$^{-1}$.
\label{maps-fig7}
}
\end{figure*}

\section{Discussion}

NGC~3079 is a unique nearby laboratory for studying the presence
of radio jets in a spiral galaxy and the effect that these might
have on the global characteristics of the galaxy.  Although little
else has been seen in the way of such well-defined radio lobes in a spiral,
 recent observations of Abell 428 have also revealed a classical
double-lobed radio source associated with a disk galaxy 
(Ledlow et al. 1998, Ledlow et al. 2001),
suggesting that NGC~3079 is no longer unique.  Our new GMRT observations
have revealed not only the details of the nuclear radio structure
(e.g. Fig. 3c) at high (3$^{\prime\prime}$) resolution, but
broad scale radio continuum emission (e.g. Fig. 1a) at low
(55$^{\prime\prime}$) resolution and much substructure.

It appears that much of the extended emission around NGC~3079 is
related to the disk.  This is seen via the envelope of emission whose
morphology follows the disk (Fig. 1a) and also from at least one feature which
forms a kpc-scale loop to the east of the northern tip 
of the galaxy (Fig. 1b) which has an HI counterpart.  Thus, the radio
continuum features are similar to those found in NGC~5775 in which
extraplanar features and supershells are detected in every component
of the ISM (Lee et al. 2001) or NGC~2613 in which a large open-topped
radio continuum arc with some associated HI spans much of the galaxy
(Chaves \& Irwin 2001).  Neither of the latter two galaxies are known
to harbour an AGN.

The nuclear radio emission and radio lobes
 have been well-delineated by our observations.  In addition, 
large radio extensions and loops are seen which may be associated with it. 
This includes the two western extensions and possibly the giant eastern
loop best seen at 615 MHz which is somewhat offset from a direct extension
of the eastern nuclear lobe. 
Fig. 8 shows the western extensions and eastern
loop in relation to the high resolution nuclear radio lobes. 
Indeed the placement of these features and their similar
sizes are suggestive of a nuclear origin.  If
the two 
large features on either side of the galaxy
 are indeed associated with the nuclear outflow
then they indicate regions of previous outflow and suggest
that the nuclear activity in NGC~3079
is episodic.  If we extrapolate a particle lifetime of $10^8$ yrs 
(Fig. 7d) and use the outer distance of the giant eastern
loop (11 kpc, Sect. 3.1.2) then the required average bulk speed
of the particles from the disk
 to this distance is $\sim$ 100 km s$^{-1}$.

The presence of offset radio continuum
and HI emission at the southern tip of the major axis argues for
some motion with respect to an IGM and this may also be affecting
the location of the eastern loop.  However, other effects may
also be important.  For example, the fact that the VLBI jet direction
is offset from the direction of the nuclear radio lobes argues
either that the source outflow direction is changing with
time or that the nuclear jets are bending through the galaxy's ISM
and being focussed along the minor axis until the outflow
moves from an ISM-dominated region to an IGM dominated region.
We speculate that the emission
around NGC~3079 may be similar to what has been observed at low
frequencies around M~87 in which the galaxy is embedded in a complex
IGM in its immediate vicinity which is due to its own nuclear
outflow (Owen et al. 2000).

\begin{figure}
\hspace{0.5in}
\psfig{figure=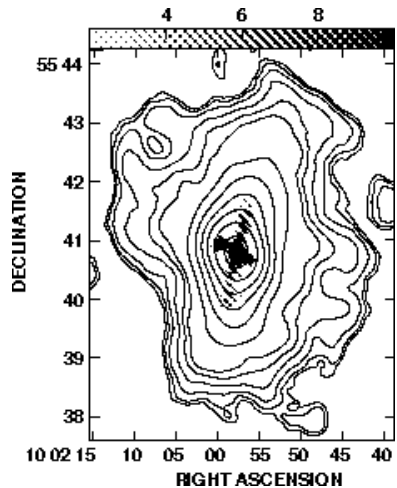,width=2.0in}
\caption{Superimposition of Fig. 2a (contours) over Fig. 3b
(greyscale).  The greyscale ranges from 2 to 10 mJy beam$^{-1}$.
\label{maps-fig8}
}
\end{figure}

\section{Conclusions}

We have observed the double-lobed spiral galaxy, NGC~3079
using the GMRT at three frequencies and detected structure
over resolutions from 3$^{\prime\prime}$ to 56$^{\prime\prime}$.
At high resolution, the structure of the nuclear radio emission is shown in
exquisite detail (Fig. 3c).  In addition, we have detected
a number of new
intermediate and broad scale features, namely:

1.  Two western extensions above and below the western
nuclear radio lobe which reach $\sim$ 10 kpc from the galaxy's
disk.  These features have associated HI emission.

2.  A giant eastern radio loop, best seen in Fig. 2a or Fig. 8,
which is slightly offset from a line extending through the
nucleus and the eastern radio lobe.  The giant loop extends
to 11 kpc from the disk.

3.  A northern loop, best seen in Fig. 1b.  There is
a large HI arc at this position.
  It is 3.8 kpc in diameter and extends
7.2 kpc from the major axis.  This loop appears to have
originated in the disk.

4.  A broad halo of emission with
a vertical scale height of at least 4.8 kpc, which follows the disk of
the galaxy.

5.  Emission at the southern tip of the major axis that is
offset from the optical major axis
suggesting that there may
be motion through broader scale IGM gas.  

No emission has been detected from either companion galaxy.

We suggest that Features 1 and 2 above may be associated with
earlier nuclear outflow  in which case the average outflow
velocity would be $\sim$ 100 km s$^{-1}$.  In addition, the
IGM that is implied in 5, above, could be a local IGM, formed
via outflow from NGC~3079 itself.    

We have also computed spectral index maps over a restricted spatial
region and have computed maps of minimum energy parameters for
this galaxy, including the cosmic ray energy density, the diffusion
length, the magnetic field strength, the particle lifetime 
(each of these representing averages along a line of sight) and finally the
power.  The total energy content is several orders of magnitude
larger than that of our Milky Way but the parameters, in general,
are typical of spiral galaxies.

\section*{Acknowledgments}

JAI wishes to thank the scientists and staff of NCRA, Pune, for graciously
allowing her to work at this institute during a sabbatical year. The GMRT 
is a national facility operated by the National Centre for Radio Astrophysics
of the Tata Institute of Fundamental Research.
This research has made use of the NASA/IPAC extragalactic database (NED)
which is operated by the Jet Propulsion Laboratory, Caltech, under contract
with the National Aeronautics and Space Administration.


\label{lastpage}

\end{document}